\documentclass[aps,showpacs,amsmath,amssymb]{revtex4}
\usepackage{bm}
\usepackage{epsfig}

\begin{document}
\title{Scheme for remote implementation of partially unknown quantum operation of two qubits in cavity QED}
\author{Liang Qiu}
\affiliation{Quantum Theory Group, Department of Modern Physics\\
University of Science and Technology of China\\ Hefei 230026,
People's Republic of China}
\author{An Min Wang}\email{anmwang@ustc.edu.cn}
\affiliation{Quantum Theory Group, Department of Modern Physics\\
University of Science and Technology of China\\ Hefei 230026,
People's Republic of China}

\begin{abstract}

By constructing the recovery operations of the protocol of remote
implementation of partially unknown quantum operation of two qubits
[An Min Wang: PRA, \textbf{74}, 032317(2006)], we present a scheme
to implement it in cavity QED. Long-lived Rydberg atoms are used as
qubits, and the interaction between the atoms and the field of
cavity is a nonresonant one. Finally, we analyze the experimental
feasibility of this scheme.

\end{abstract}


\maketitle

\section{Introduction}\label{sec1}

The remote implementation of quantum operation (RIO) is defined as
that a quantum operation performed on the local system(Alice's
one) is teleported and acts on an unknown state belonging to the
remote system (Bob's) \cite{Huelga,Huelga1,Huelga2,Wang}. In
Ref.\cite{Huelga} and subsequent research\cite{Huelga2}, the
authors conclude that, the optimal LOCC(local operation and
classical communication) procedure  to implement remotely an
arbitrary unitary operator $U$ on a qudit with the shared
entanglement is by the means of ``bidirectional state
teleportation". Furthermore, the remote implementation of a
unitary transformation on the state of a qubit is
studied\cite{Huelga1}. Just as the teleportation \cite{Bennett} of
an unknown quantum state, in the process of RIO, entangled states
are used. However, the cost of entanglement resources is dependent
on whether quantum operations are unknown or partially unknown
(known).  When it comes to the ``partially unknown operation'', it
implies that the quantum operation satisfies some given restricted
conditions. As in the reference \cite{Huelga1}, the authors
considere the case of two kinds of one-qubit operations, one of
them  only has non-zero diagonal elements:  arbitrary rotations
around a fixed direction $\overrightarrow{n}$, the other just has
non-zero offdiagonal elements: it a $\pi$ rotation about an
arbitrary direction lying in a plane orthogonal to
$\overrightarrow{n}$. For the cases more than one qubit, for
example, N qubits case, Wang \cite{Wang} proves that the quantum
operations only with one non-zero element in every row and every
column of their representation matrices can be able to be
implemented remotely in a faithful and determined way, if we only
have N $e$-bits and use Hadamard gates to transfer the effect of
operation to Bob's qubits. In this paper, our motivation is just
to present a scheme of remote implementation of partially unknown
quantum operations of two qubits based on the well-known
technology and method. Recently, by using a linear optics set-up,
a remote rotation by $120^{\circ}$ about the $z$ axis has been
implemented experimentally on the target photons\cite{Xiang}.
Moreover, the authors claim that the scheme can be generalized to
implement the single qubit subsets discussed in\cite{Huelga1}.

Cavity QED \cite{Turchette,Zheng}, optical systems
\cite{Bouwmeester}, ion trap \cite{Monroe} and NMR
\cite{Gershenfeld} are all used to demonstrate quantum information
processing and quantum computation. Recently, cavity QED
technology has attracted  a lot of interest. In this context,
cavity QED with circular Rydberg atoms and superconducting
cavities presents a peculiar interest. In cavity QED, quantum
logic gates are constructed \cite{Zheng}; Bell-state
\cite{Solano,Ye,Chen}, GHZ state(W state) \cite{Guo}, even the
$n$-particle entangled state \cite{Zheng1} are generated. Some
important tasks of quantum information processing and quantum
communication, such as teleportation \cite{Jin,Pires}, quantum
state sharing \cite{Xue} and Grover's search (\cite{Yamaguchi} and
the references in) are successfully implemented by using cavity
QED. Some of experimental demonstrations of quantum information
and quantum computation in cavity QED have also been proposed
\cite{Osnaghi,Riebe}.

In this paper, by constructing the recovery operations of remote
implementation of two-qubit partially unknown quantum operations,
 we find it is possible to carry out the quantum
information processing in cavity QED. Then we present the scheme
in cavity QED and analyze its experimental feasibility. The
remainder of this paper is organized as follows: In
Sec.\ref{sec2}, a simple introduction of Wang's protocol is
presented and the recovery operations are constructed; in
Sec.\ref{sec3}, the scheme of remote implementation of two-qubit
partially unknown quantum operation in cavity QED is presented;
finally, in Sec.\ref{sec4}, we present the discussion and
conclusion.

\section{Remote implementation of two-qubit partially unknown quantum
operations}\label{sec2}

The structure of the two-qubit partially unknown quantum
operations which can be remotely implemented and the recovery
operation are presented in Ref.\cite{Wang}. Because there is only
one nonzero element in every row and every column of the matrices
of the two-qubit operations, there are 24 kinds of operations.
They can be written as:
\begin{equation}
T_{2}(x,t)=\sum\limits_{m=00}^{11}t_{m}|m\rangle\langle
p_{m}(x)|=\left(
\begin{array}{cccc}
t_{00}&0&0&0\\
0&t_{01}&0&0\\
0&0&t_{10}&0\\
0&0&0&t_{11}\\
\end{array}
\right)R_{2}(x),
\end{equation}
where $p_{m}(x)$ is the corresponding elements of $p(x)$, which is
the permutations of the list $(00, 01, 10, 11)$ and $x=1, 2, \cdots,
24$. So $p(x)=(p_{00}(x), p_{01}(x), p_{10}(x), p_{11}(x))=(00, 01,
10, 11), (00, 01, 11, 10), \cdots, (11, 10, 01, 00)$. A part of the
recovery operation:
\begin{equation}
R_{2}(x)=T_{2}(x,0)=\sum\limits_{m=00}^{11}|m\rangle\langle
p_{m}(x)|,
\end{equation}
as the above $x=1, 2, \cdots, 24$. The recovery operations
correspond to the two-qubit quantum operations by Alice sends the
$x$ to Bob with 5 bits.

We briefly recall the remote implementation of two-qubit partially
unknown quantum operation \cite{Wang}. Two entangled states
$|\Psi^{+}\rangle_{A_{1}B_{1}}, |\Psi^{+}\rangle_{A_{2}B_{2}}$
work as the channel, qubits $A_{1}, A_{2}$ belong to Alice and
$B_{1}, B_{2}$ belong to Bob. Another two qubits $Y_{1}, Y_{2}$ in
an unknown state $|\xi\rangle_{Y_{1}Y_{2}}$ belong to Bob too.
Two-qubit partially unknown quantum operations acted by Alice can
work on qubits $B_{1}, B_{2}$ as follows. Bob first performs two
controlled-{\sc not} ($C_{\rm not}$) transformations by using
$Y_{1}, Y_{2}$ as control qubits and $B_{1}, B_{2}$ as target
qubits, respectively. Then he measures his qubits $B_{1}$ and
$B_{2}$ in the computational basis $|b_{1}\rangle_{B_{1}}\langle
b_{1}|\otimes|b_{2}\rangle_{B_{2}}\langle b_{2}|$, where $b_{1},
b_{2}=0, 1$ and sends the results to Alice. After receiving the
two bits, Alice carries out the quantum operations
$\sigma_{b_{1}}^{A_{1}}\otimes\sigma_{b_{2}}^{A_{2}}$ on her two
qubits $A_{1}, A_{2}$ respectively. Subsequently, she acts
$T_{2}(x,t)$ on $A_{1}A_{2}$ and executes two Hadamard gate
transformation $H_{A_{1}}\otimes H_{A_{2}}$. In the end, she
measures her two qubits in the basis $|a_{1}\rangle_{A_{1}}\langle
a_{1}|\otimes|a_{2}\rangle_{A_{2}}\langle a_{2}|(a_{1}, a_{2}=0,
1)$ and sends the results and $x$ to Bob. As  having been
mentioned, the transmission of $x$ is to let Bob choose $R_{2}(x)$
correctly. With this information, Bob's recovery operations are
\begin{equation}
\mathcal{R}(a_{1}, a_{2},
x)=\left\{[(1-a_{1})\sigma_{0}+(a_{1}\sigma_{3})]_{Y_{1}}
\otimes[(1-a_{2})\sigma_{0}+(a_{2}\sigma_{3})]_{Y_{2}}\right\}
R_{2}(x)_{Y_{1}Y_{2}}.
\end{equation}
With this steps, the two-qubit partially unknown operations can be
remotely implemented on qubits $Y_{1}Y_{2}$.

The possible obstacle to demonstrate the protocol is the recovered
operation $R_{2}(x)$. Fortunately, we can construct $R_{2}(x)$ by
using the $C_{\rm not}$ gate  and the {\sc not} gate $\sigma_{x}$.
Actually, this comes from the fact that any multiqubit logic gate
can be decomposed as $C_{\rm not}$ transformations and single
qubit logic gates \cite{Nielsen}. We have:
\begin{eqnarray}
R_{2}(1)&=&I\otimes I,\\
R_{2}(2)&=&C_{\rm not}(Y_{1},Y_{2}),\\
R_{2}(3)&=&C_{\rm not}(Y_{2},Y_{1}) C_{\sc not}(Y_{1},Y_{2})
C_{\rm not}(Y_{2},Y_{1}),\\
R_{2}(4)&=&C_{\rm not}(Y_{2},Y_{1})
C_{\rm not}(Y_{1},Y_{2}),\\
R_{2}(5)&=&C_{\rm not}(Y_{1},Y_{2}) C_{\rm not}(Y_{2},Y_{1}),\\
R_{2}(6)&=&C_{\rm not}(Y_{2},Y_{1}),\\
R_{2}(7)&=&C_{\rm not}(Y_{1},Y_{2})(I\otimes\sigma_{1}),\\
R_{2}(8)&=&(I\otimes\sigma_{1}),\\
R_{2}(9)&=&(\sigma_{1}\otimes I) C_{\rm not}(Y_{1},Y_{2})
C_{\rm not}(Y_{2},Y_{1}),\\
R_{2}(10)&=&C_{\rm not}(Y_{2},Y_{1})(I\otimes\sigma_{1}),\\
R_{2}(11)&=&C_{\rm not}(Y_{2},Y_{1})(\sigma_{1}\otimes I)
C_{\rm not}(Y_{1},Y_{2}) C_{\rm not}(Y_{2},Y_{1}),\\
R_{2}(12)&=&C_{\rm not}(Y_{2},Y_{1}) C_{\rm
not}(Y_{1},Y_{2})(I\otimes\sigma_{1}),\\
R_{2}(13)&=&C_{\rm
not}(Y_{2},Y_{1}) C_{\rm
not}(Y_{1},Y_{2})(\sigma_{1}\otimes I),\\
R_{2}(14)&=&C_{\rm not}(Y_{2},Y_{1}) C_{\rm
not}(Y_{1},Y_{2})(\sigma_{1}\otimes I)
 C_{\rm not}(Y_{2},Y_{1}),\end{eqnarray}

\begin{eqnarray}
R_{2}(15)&=&C_{\rm not}(Y_{2},Y_{1})(\sigma_{1}\otimes I),\\
R_{2}(16)&=&C_{\rm not}(Y_{1},Y_{2})(\sigma_{1}\otimes I) C_{\rm
not}(Y_{2},Y_{1}),\\
R_{2}(17)&=&(\sigma_{1}\otimes I),\\
R_{2}(18)&=&C_{\rm not}(Y_{1},Y_{2}) (\sigma_{1}\otimes I),\\
R_{2}(19)&=&(I\otimes\sigma_{1})
C_{\rm not}(Y_{2},Y_{1}),\\
R_{2}(20)&=&C_{\rm not}(Y_{1},Y_{2})(I\otimes \sigma_{1})
 C_{\rm not}(Y_{2},Y_{1}),\\
R_{2}(21)&=&C_{\rm not}(Y_{2},Y_{1}) (\sigma_{1}\otimes I)
C_{\rm not}(Y_{1},Y_{2}),\\
R_{2}(22)&=&C_{\rm not}(Y_{2},Y_{1}) C_{\rm not}(Y_{1},Y_{2})
(I\otimes\sigma_{1}) C_{\rm not}(Y_{2},Y_{1}),\\
R_{2}(23)&=&(\sigma_{1}\otimes I) C_{\rm not}(Y_{1},Y_{2}),\\
R_{2}(24)&=&(\sigma_{1}\otimes\sigma_{1}).
\end{eqnarray}
where $C_{\rm not}(Y_{1},Y_{2})$ means that we use qubit $Y_{1}$
as the control qubit, $Y_{2}$ as the target qubit to do the
controlled-{\sc not} transformation, and $C_{\rm
not}(Y_{2},Y_{1})$ means we use qubit $Y_{2}$ as the control qubit
and  qubit $Y_{1}$ as the target qubit.  In addition, $\sigma_{i}$
is the Pauli matrices, with $\sigma_1=\left(
\begin{array}{cc}
0&1\\
1&0\\
\end{array}
\right), \sigma_2=\left(
\begin{array}{cc}
0&-i\\
i&0\\
\end{array}
\right), \sigma_3=\left(
\begin{array}{cc}
1&0\\
0&-1\\
\end{array}
\right)$ and $I$ is the $2\times2$ identity matrix.

\section{Demonstrate the protocol in cavity QED}\label{sec3}

In the original protocol\cite{Wang}, after constructing the
recovery operations, all the operations to remotely implement
two-qubit partially unknown quantum operations are two-qubit
$C_{\rm not}$ gate, single qubit logic gates: Hadamard gate and
pauli operations. The quantum resources we need is just two
$e$-bits. All of them can be realized by using cavity QED.

As we have shown in the introduction, the preparation of two-qubit
entangled states has been demonstrated in many
papers\cite{Solano,Ye,Chen}. In the reference \cite{Zheng}, Zheng
and Guo proposed a realizable scheme of two-atom controlled-{\sc
not} gate in cavity QED.
 In the protocol, ladder-type three-level(denoted by
$|g\rangle, |e\rangle$ and $|i\rangle$) atoms are used. In order
to make sure that $|i\rangle$ is not affected by the atom-cavity
interaction, the transition frequency between the state
$|e\rangle$ and $|i\rangle$ is highly detunned from the cavity
frequency.

Let us start with considering two identical ladder-type
three-level atoms simultaneously interacting with a single cavity.
There is no energy exchange between the atomic system and the
cavity under the approximation $\delta\gg g$. In the case of the
cavity field in the vacuum state, the effective Hamiltonian is
given by:
\begin{equation}
\label{nonrcavity}
H=\lambda\left[\sum\limits_{j=1,2}|e_{j}\rangle\langle
e_{j}|+(S_{1}^{+}S_{2}^{-}+S_{1}^{-}S^{+}_{2})\right],
\end{equation}
where $\lambda=g^{2}/\delta, S_{j}^{+}=|e_{j}\rangle\langle
g_{j}|$ and $S_{j}^{-}=|g_{j}\rangle\langle e_{j}|$, with
$|g_{j}\rangle, |e_{j}\rangle(j=1, 2)$ being the ground and
excited states of the atom. $a^{+}, a$ are the creation and
annihilation operators of the cavity mode. $g$ is  the atom-cavity
coupling strength, and $\delta$ is the detuning  between the
atomic transition frequency $\omega_{0}$ and cavity frequency
$\omega$.

Now, the $C_{\rm not}$ gate can be realized as follows\cite{Zheng}:
atom 2 passes through classical field tuned to the transitions
$|g\rangle\rightarrow|e\rangle$ and $|e\rangle\rightarrow|i\rangle$,
the amplitudes and phases of which are appropriately chosen, so we
have:
\begin{equation}
\label{classical fields 1}
\begin{array}{l}
|e_{2}\rangle\rightarrow\frac{1}{\sqrt{2}}\left(|e_{2}\rangle+|g_2\rangle\right)
\rightarrow\frac{1}{\sqrt{2}}\left(|i_{2}\rangle+|g_{2}\rangle\right),\\
|g_{2}\rangle\rightarrow\frac{1}{\sqrt{2}}\left(|g_{2}\rangle-|e_{2}\rangle\right)
\rightarrow\frac{1}{\sqrt{2}}\left(|g_{2}\rangle-|i_{2}\rangle\right),\\
\end{array}
\end{equation}
after this, we send the atom 1 and 2 into the nonresonant cavity
Eq.(\ref{nonrcavity}) simultaneously, by choosing the interaction
time $\lambda t=\pi$:
\begin{equation}
\begin{array}{l}
|g_{1}g_{2}\rangle\rightarrow|g_{1}g_{2}\rangle,\\
|g_{1}i_{2}\rangle\rightarrow|g_{1}i_{2}\rangle,\\
|e_{1}g_{2}\rangle\rightarrow|e_{1}g_{2}\rangle,\\
|e_{1}i_{2}\rangle\rightarrow-|e_{1}i_{2}\rangle,\\
\end{array}
\end{equation}
then atom 2 passes through two classical fields tuned to the
transitions $|e\rangle\rightarrow|i\rangle$ and
$|g\rangle\rightarrow|e\rangle$ respectively by appropriately
choosing the amplitudes and phases of the classical fields:
\begin{equation}
\label{classical fields 2}
\begin{array}{l}
|g_{2}\rangle\rightarrow\frac{1}{\sqrt{2}}(|g_{2}\rangle+|e_{2}\rangle),\\
|i_{2}\rangle\rightarrow|e_{2}\rangle\rightarrow\frac{1}{\sqrt{2}}(|e_{2}\rangle-|g_{2}),\\
\end{array}
\end{equation}
with this steps, we will obtain the $C_{\rm not}$ transformation:
\begin{equation}
\begin{array}{l}
|g_{1}g_{2}\rangle\rightarrow|g_{1}g_{2}\rangle,\\
|g_{1}e_{2}\rangle\rightarrow|g_{1}e_{2}\rangle,\\
|e_{1}g_{2}\rangle\rightarrow|e_{1}e_{2}\rangle,\\
|e_{1}e_{2}\rangle\rightarrow|e_{1}g_{2}\rangle,\\
\end{array}
\end{equation}

In any physical system, single qubit gates are easily performed.
To the atoms, these single qubit gates can be realized by using
rotations \cite{Yamaguchi}.  We could also realize Hadamard gate
in cavity QED by considering an atom through an initially empty
resonant cavity. In the interaction picture, the Hamiltonian is
\cite{Gor,Cao,Cardoso}:
\begin{equation}
H_{I}=g\left[a^{+}S^{-}+aS^{+}\right].
\end{equation}
It's the Jaynes-Cummings model, where $a^{+}, a$ are the creation
and annihilation operator of the cavity field.
$S^{+}=|e\rangle\langle g|, S^{-}=|g\rangle\langle e|$, with
$|g\rangle, |e\rangle$ being the ground and excited states of the
atom.

The Hadamard gate can be realized as follows: at first we send the
atom through the initially empty cavity and choose the interaction
time $gt=\pi$, after that we let atom cross the classical field
 $R_{+}$\cite{Serra}. The process is:
\begin{eqnarray}
|g\rangle&\stackrel{JC}{\longrightarrow}&|g\rangle
\stackrel{R_{+}}{\longrightarrow}\frac{1}{\sqrt{2}}(|g\rangle+|e\rangle),\\
|e\rangle&\stackrel{JC}{\longrightarrow}&-|e\rangle
\stackrel{R_{+}}{\longrightarrow}\frac{1}{\sqrt{2}}(|g\rangle-|e\rangle),
\end{eqnarray}
where $R_{+}$ represents the action of the Ramsey zone
\begin{equation}
R_{+}=\frac{1}{\sqrt{2}}(I+i\sigma_{y}).
\end{equation} Thus, in cavity QED, not only the $C_{\rm not}$ operation, but also the
Hadamard gate has been realized.

Now we present the scheme to implement the remote implementation
of two-qubit partially unknown quantum operations in cavity QED
with three steps, where the logical states $|1\rangle$ and
$|0\rangle$ are represented by atom state $|e\rangle$ and
$|g\rangle$, if we use Rydberg atoms as the qubits, and the shared
two Bell pairs are
\begin{equation}
\frac{1}{2}\left(|gg\rangle_{A_1B_1}+|ee\rangle_{A_1B_1}\right)
\otimes\left(|gg\rangle_{A_2B_2}+|ee\rangle_{A_2B_2}\right),
\end{equation}
and the unknown state of two qubits is:
\begin{equation}
|\xi\rangle_{Y_{1}Y_{2}}=y_{gg}|gg\rangle+y_{ge}|ge\rangle+y_{eg}|eg\rangle+y_{ee}|ee\rangle.
\end{equation}

{\em Step 1}: let atom $B_{1}$ cross two classical fields tuned to
the transitions $|g\rangle\rightarrow|e\rangle$ and
$|e\rangle\rightarrow|i\rangle$ given by Eq.(\ref{classical fields
1}), respectively. Then, atoms $Y_{1}, B_{1}$ are sent into the
nonresonant cavity simultaneously given by Eq.(\ref{nonrcavity}).
After they pass through the cavity, atom $B_{1}$ crosses two
classical fields tuned to the transitions
$|e\rangle\rightarrow|i\rangle$ and
$|g\rangle\rightarrow|e\rangle$ shown in Eq.(\ref{classical fields
2})($C_{\rm not}(Y_{1},B_{1})$). At the same time, to the atom
$B_{2}$ and $Y_{2}$, we operate them by correspondingly replacing
$B_{1}, Y_{1}$ with $B_{2}, Y_{2}$($C_{\rm not}(Y_{2},B_{2})$).
Then measuring two qubits $B_{1}, B_{2}$ in the basis
$|g\rangle_{B_{1(2)}}, |e\rangle_{B_{1(2)}}$ with the result
$b_{1}, b_{2}$, and assuming $b_{1}=b_{2}=0$, which means we get
$|g\rangle_{B_{1}}, |g\rangle_{B_{2}}$,  the state of the system
becomes
\begin{equation}
\left(
y_{gg}|gggg\rangle+y_{ge}|gege\rangle+y_{eg}|egeg\rangle+y_{ee}|eeee\rangle
\right)_{A_{1}A_{2}Y_{1}Y_{2}}\otimes|gg\rangle_{B_{1}B_{2}}.
\end{equation}

{\em Step 2}: With $b_{1}, b_{2}$, we do the
rotations\cite{Jin,Yamaguchi} on $A_{1}, A_{2}$ so as to realize
$\sigma_{b_{1}}^{A_{1}}$ and $\sigma_{b_{2}}^{A_{2}}$.  Following,
we act $T_{2}(x,t)$ on $A_{1}A_{2}$. After doing the rotations on
atoms $A_{1}, A_{2}$ to realize the $H_{A_{1}}, H_{A_{2}}$, two
atoms $A_{1}, A_{2}$ are measured in the basis
$|g\rangle_{A_{1(2)}}, |e\rangle_{A_{1(2)}}$.  External classical
microwave resources resonant on the $|e\rangle - |g\rangle$
produce these rotations on the two atoms respectively. The
amplitude and phase of these sources are carefully tunned to
produce the required transitions. For example, if we want to
remotely implement two-qubit partially unknown quantum operation
\begin{equation}
T_{2}(10,t)=\sum\limits_{m=gg}^{ee}t_{m}|m\rangle\langle
p_{m}(10)|= \left(
\begin{array}{cccc}
0&t_{gg}&0&0\\
0&0&t_{ge}&0\\
0&0&0&t_{eg}\\
t_{ee}&0&0&0\\
\end{array}
\right),
\end{equation}
$p_{m}(10)$ is the corresponding element of $p(10)=(ge, eg,
ee,gg)$. We assume that measurement outputs of atoms $A_{1}, A_{2}
$ are $a_{1}=a_{2}=1$, which tell us that the atoms are in the
state $|e\rangle_{A_{1}}, |e\rangle_{A_{2}}$. After  Step 2, we
will obtain
\begin{equation}
\left(y_{gg}t_{ee}|gg\rangle+y_{ge}t_{gg}|ge\rangle-y_{eg}t_{ge}|eg\rangle-y_{ee}t_{eg}|ee\rangle\right)_{Y_{1}Y_{2}}
\otimes|ee\rangle_{A_{1}A_{2}}\otimes|gg\rangle_{B_{1}B_{2}}.
\end{equation}
 Now let us focus on the recovery operation.

{\em Step 3}: With $a_{1}=a_{2}=1$ and $x=01010$(which is used to
denote decimal system 10), for the atoms $Y_{1}, Y_{2}$, after
doing the rotation to realize $\sigma_{1}^{Y_{2}}$, we perform
them just same as the step 1 by correspondingly substituting
$Y_{1}, B_{1}$ with $Y_{2}, Y_{1}$($C_{\rm not}(Y_{2},Y_{1})$).
Then we do rotations to implement $\sigma_{3}^{Y_{1}},
\sigma_{3}^{Y_{2}}$. Thus the system evolves into
\begin{equation}
T_{2}(10,t)\left(y_{gg}|gg\rangle+y_{ge}|ge\rangle+y_{eg}|eg\rangle+y_{ee}|ee\rangle\right)_{Y_{1}Y_{2}}
\otimes|ee\rangle_{A_{1}A_{2}}\otimes|gg\rangle_{B_{1}B_{2}}.
\end{equation}
With the three steps, we can remotely implement the $T_{2}(10,t)$
on the atoms $Y_{1}, Y_{2}$. The simple figure of the experimental
apparatus is shown in Fig.1.

\begin{figure}[h]
\begin{center}
\includegraphics[scale=0.8]{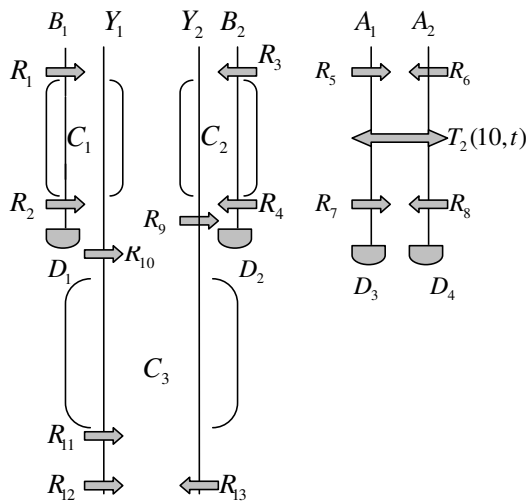}
\end{center}
\caption{Experimental apparatus. $R_{i}(i=1,2,\cdots,13)$ is the
appropriately chosen classical field to realize the transitions
among atomic  levels, $Hadamard$ gates and pauli operations.
$C_{i}(i=1,2,3)$ is the nonresonant cavity, and the two atoms must
be sent into it simultaneously. $D_{i}(i=1,2,3,4)$ is the
measurement we do on atom in the basis $|g\rangle,
|e\rangle$.}\label{mypic2}
\end{figure}

\section{Discussion and conclusion}\label{sec4}

We consider the experimental realization of our protocol.  On the
one hand, we consider the radiation of the atom. To the $C_{\rm
not}$ operation, if choosing $\delta=10g$ and $g=2\pi\times24$kHz
\cite{Zheng,Brune}, the interaction time of cavity-field is
$\pi\delta/g^{2}\approx2\times10^{-4}$s. The time needed for
the atom tuned with classical field is on the order of
$6.3\times10^{-6}s$ \cite{Pires}, thus it is negligible at this
scale\cite{Yamaguchi}. So the time needed to implement the scheme
is on the order of $10^{-3}$s, much shorter than the radiative
time of the Rydberg atom with principal quantum numbers $49, 50$
and $51$, which is about $3\times10^{-2}$s.  On the other hand, we
consider the cavity decay. With present cavity technology, a
cavity with a quality factor $Q=10^{8}$ is experimentally
achievable\cite{Brune}. As discussed in \cite{Zheng}, the photon
lifetime in the cavity whose cavity frequency is about 50 GHz is
$\tau_{C}=\frac{Q}{2\pi\nu}\approx3\times10^{-4}$s. In the present
protocol, that the cavity is always in the vacuum state result in
the suppressed cavity decay. Therefore, the cavity has only about
0.01 probability of being excited during the passage of the atoms
through the cavity and the efficient decay time of the cavity is
about $3\times10^{-2}$s, on the order of the atomic radiative
time, which is much longer than the time needed to implement the
scheme. In 2001, an experiment of preparing EPR pairs with two
atoms using the present model has been implemented\cite{Osnaghi}.
To the resonant cavity, which we use it to realize Hadamard
transformation, the atom acting as the qubit must have a
sufficiently long excited life-time. Luckily, the Rydberg atom
with principal quantum numbers 50 and 51 is a good candidate,
because the interaction time( with the same characters, we have
the interaction time $\pi/g=2\times10^{-5}s$) is much shorter than
the atomic radiative time. Hence the time to complete the remote
implementation of two-qubit partially unknown quantum operations
is much shorter than that of atom radiation.  However, there is
still a difficulty to carry out our protocol: distinguishing the
two atoms after they flying out of the cavity. Fortunately, we can
use the method proposed in\cite{Riebe}, which has developed a
technique to address any specified target ion using tightly
focused laser beams and by changing their internal states to
``hide" the remaining ions from the target ion's fluorescence so
that they are insensitive to the fluorescent light. Therefore, our
scheme is realizable based on current cavity technology.

In order to realize the $C_{\rm not}$ operation, we require the
two atoms are sent through a cavity simultaneously. Hence we would
like to estimate the influence when the simultaneous is not
exactly satisfied, that is we estimate the fidelity between the
result that one atom enters the cavity sooner than the other by
$0.01t$ and the Eq.(40) through an numerical calculation, and get
the fidelity as high as $99.8\%$. The calculation results are
showed in Fig.2. Therefore, our protocol are slightly affected.
\begin{figure}[h]
\begin{center}
\includegraphics[scale=0.8]{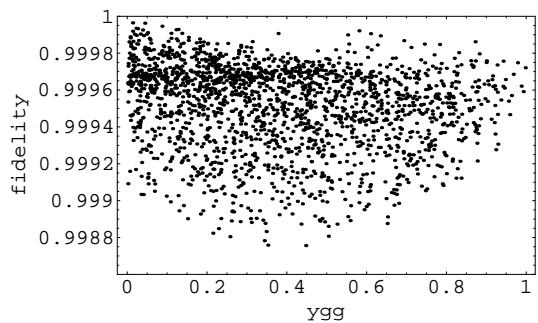}
\end{center}
\caption{Fidelity vs the value of $ygg$. In order to get the
numerical results, we have assumed
$t_{gg}=t_{ge}=t_{eg}=t_{ee}=x$, where $|x|^2=1$ resulting from
the fact that $T_{2}(10,x)$ is a unitary transformation. In the
calculation, $y_{gg}, y_{ge}, y_{eg}, as well as y_{ee}$ are all
arbitrary positive real numbers  and the previous three run their
values. We make sure $y_{gg}^2+y_{ge}^2+y_{eg}^2\leq1$,
$y_{gg}^2+y_{ge}^2+y_{eg}^2+y_{ee}^2=1$.}\label{mypic2}
\end{figure}

In summary, we propose the scheme to remotely implement two-qubit
partially unknown quantum operation in cavity QED. In order to do
this, the recovery operations are constructed by $C_{\rm not}$ and
single-qubit pauli operations, thus all the operations are just
$C_{\rm not}$, Hadamard gate and pauli operations. We realize them
and draw the conclusion that the scheme can be demonstrated in
cavity QED. Besides, from the Eq.(1), the two-qubit partially
unknown quantum operations can be constructed in cavity QED
because we have realized $R_{2}(x)$.

We are grateful all the collaborators of our quantum theory group in
the Institute for Theoretical Physics of our university. This work
was funded by the National Fundamental Research Program of China
under No. 2001CB309310, and partially supported by the National
Natural Science Foundation of China under Grant No. 60573008.

\end{document}